\documentclass[aps,showpacs, showkeys, 12pt]{revtex4}

\usepackage{amsfonts}
\usepackage{graphicx}
\usepackage{amsmath}
\usepackage{amssymb}
\usepackage{latexsym}
\usepackage{psfrag}
\usepackage{anysize}

\newcommand{\kv}{\mathbf{k}}
\newcommand{\qv}{\mathbf{q}}

\begin{document}


\title{\textbf{Superconductivity in Graphene-Lithium}}

\author{D. M. Guzman}
\email{guzmand@purdue.edu}
\affiliation{School of Materials Engineering, Purdue University, West Lafayette, Indiana 47907, USA}

\author{H. M. Alyahyaei}
\email{halya001@ucr.edu}
\affiliation{Department of Materials Science and Engineering, University of California, Riverside, California 92521, USA}

\author{R. A. Jishi}
\email{rjishi@calstatela.edu}
\thanks{Corresponding author}
\affiliation{Department of Physics and Astronomy, California State University, Los Angeles, California 90032, USA}

\begin{abstract}
We present first-principles calculations on systems consisting of a few layers of graphene and lithium. 
In particular, we investigate the evolution of the electron-phonon coupling strength with 
increasing number of layers. We find that, for intercalated systems such as 
C$_6$-Li-C$_6$ or C$_6$-Li-C$_6$-Li-C$_6$, the electron-phonon coupling is weak. 
However, for systems of equal number of graphene and lithium layers, such as 
C$_6$-Li or C$_6$-Li-C$_6$-Li, the electron-phonon coupling is strong. 
We investigate the optimal configuration that yields the highest superconducting transition temperature.
\end{abstract}


\maketitle

\section{Introduction}
\label{sec:intro}
Ever since its discovery, graphene\cite{Novoselov1} has shown a great potential for applications 
in nano-technology\cite{Katsnelson, Novoselov2, Zhang, Nair}, especially in electronic 
transport\cite{elec-graphene}. Studies of superconductivity in graphite intercalation 
compounds (GICs)\cite{Hannay, Koike, Kobayashi, Alexander, Iye, Al-Jishi0, Emery1, Weller, Emery2, Walters}, 
recent experimental evidence of a superconducting state in graphite-sulfur 
composites\cite{daSilva, Moehlecke}, and the growing interest in superconducting 
nano-devices\cite{qd, set, jj} have motivated theoretical investigation of possible 
superconductivity in the two-dimensional graphene 
systems\cite{Uchoa, Zhao, Black, Honerkamp, Kopnin, Gonzalez, Loktev}. 
Single- and few-layer graphene may exhibit a superconducting state upon adatom deposition or intercalation. 
As in the case of alkali-metal and alkaline-earth GICs, superconductivity in modified-graphene systems can 
be explained by the electron-phonon coupling enhancement that arises from the presence of an 
intercalant-derived band as well as graphitic $\pi$-bands at the Fermi level\cite{Al-Jishi0, Al-Jishi1, Al-Jishi2}.

Theoretical studies within the density functional theory (DFT) framework were conducted on 
single- and few-layer graphene systems with intercalated and deposited dopants. 
The electron-phonon coupling parameter $\lambda$ was found to be $0.60$ and $0.80$, 
respectively, in calcium intercalated graphene bilayers and trilayers. 
The superconducting transition temperature T$_c$ was estimated to be $4.1$ K for 
Ca-intercalated bilayer and $10.1$ K for Ca-intercalated trilayer\cite{Jishi3}. 
Recently, the possibility of a superconducting state in lithium covered graphene monolayers 
was discussed by Profeta {\it et al.} \cite{Lidepo}. They argued that removal of quantum 
confinement, in going from LiC$_6$ GIC to Li-covered monolayer graphene, could bring the Li-derived 
states down in energy to the Fermi level. This enhances the electron-phonon coupling strength, 
making Li-covered single-layer graphene a superconductor at T$_{c}=8.1$ K.

The recent fabrication of lithium intercalated graphene bilayer \cite{sugawara} raises the question 
of whether this two-dimensional material would also exhibit a superconducting state. In addition, 
the possibility of a further enhancement of the electron-phonon coupling strength in few-layer 
graphene systems intercalated with lithium, remains an open issue. 

In this work, we study the effect of increasing number of Li-doped graphene layers on the 
electron-phonon coupling strength via first-principles density functional theory. 
Details of our methods are presented in Section \ref{sec:meths}. The electronic structure of 
lithium covered graphene and lithium intercalated graphene layers is calculated. 
It is found that only Li-covered graphene systems exhibit a coexistence of 
Li-derived $s$-bands and graphitic $\pi$-bands at the Fermi level. In the case of 
Li-intercalated graphene systems, the interlayer band (derived from Li states) is unoccupied. 
From the vibrational modes and electron-phonon coupling calculations we find that a significant 
coupling of electrons to out-of-plane vibrations is necessary to achieve a significant 
enhancement of the total electron-phonon coupling strength. 

In  Section \ref{sec:systems} we describe the crystal structure of the Li-doped graphene 
systems investigated in this work, and in Section \ref{sec:meths} details of the methods 
of calculation are presented.
The results of the electronic structure and vibrational modes calculations are presented 
and discussed in Section \ref{sec:elec} and \ref{sec:mode}, respectively. 
Conclusions are given in Section \ref{sec:con}.


\section{Crystal Structure}
\label{sec:systems}
Systems consisting of equal numbers of graphene and lithium layers (Li-covered graphene) 
and lithium intercalated graphene layers were systematically analyzed from first-principles. 
Calculations of the electronic structure and vibrational modes were carried out on systems 
of up to four lithium and four graphene layers for Li-covered graphene and up to two 
lithium and three graphene layers for Li-intercalated graphene layers. In all the cases 
considered, the unit cell is a $(\sqrt{3}a\times\sqrt{3}a)\ R\ 30^\circ$ supercell of 
the graphene sheet unit cell. In agreement with the stage-1 Li-GIC, the carbon layer 
stacking is taken to be $AAAA$. The separation between adjacent carbon and Li 
layers is set at $1.855$ \AA, similar to the corresponding value in Li GICs.
The in-plane lattice constants are $a=b=4.33$ \AA. Since the DFT calculations are 
carried out on a 3-dimensional crystal, the $c$-lattice constant is taken to be 
sufficiently large so as to render the system, in effect, two-dimensional.

\begin{figure}[ht]
    \includegraphics{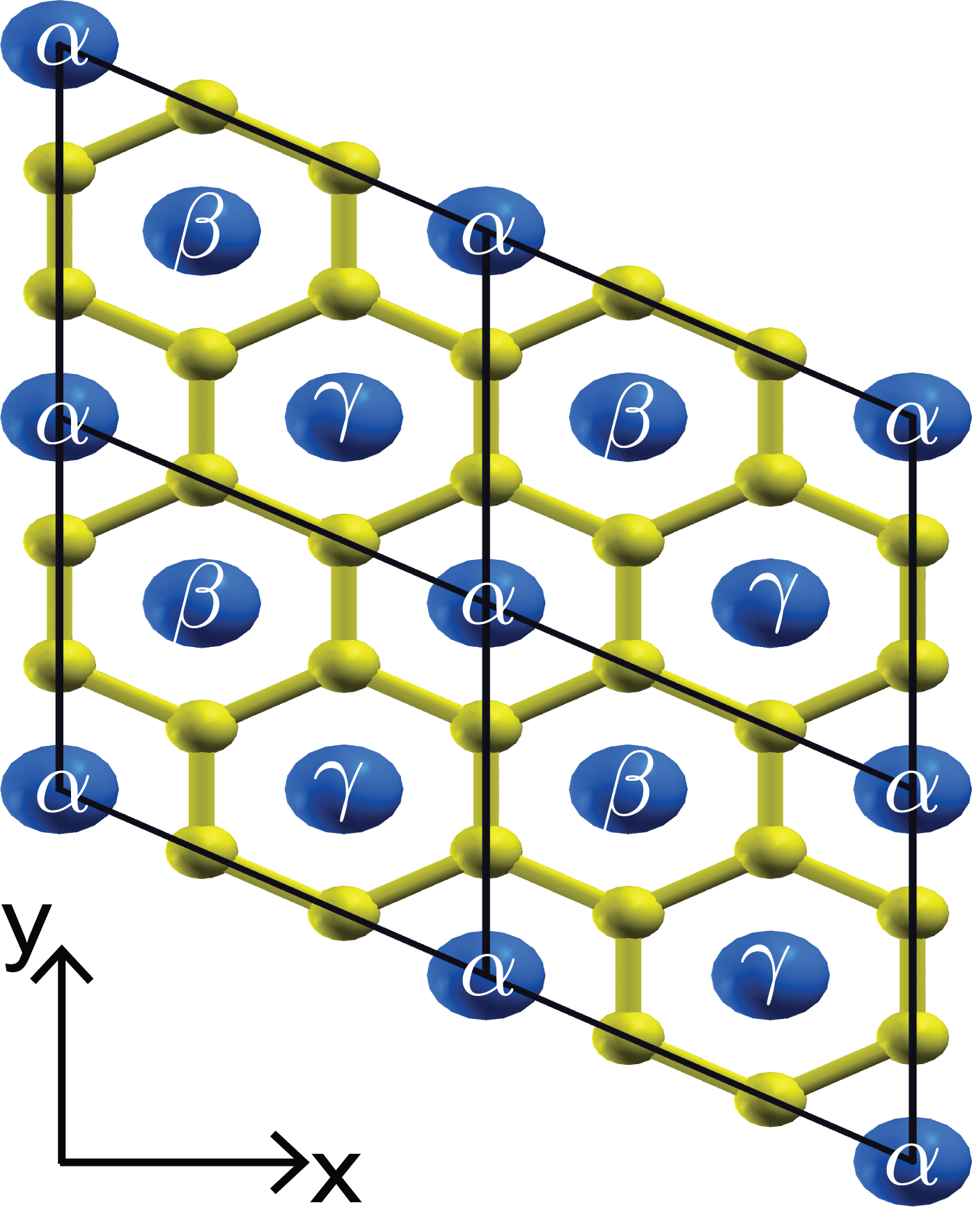}
\caption{(Color online) The graphene sites labeled $\alpha$, $\beta$, and $\gamma$. In the 
Li-covered system denoted by C$_6$Li$_{\alpha\beta\gamma}$, Li atoms sit over the $\alpha$-sites
of the first graphene sheet, over the $\beta$-sites of the second graphene sheet, and over the 
$\gamma$-sites of the third graphene sheet.}
\label{fig:abc}
\end{figure}

For systems of the form C$_6$-Li-C$_6$-Li and C$_6$-Li-C$_6$-Li-C$_6$, two stacking 
configuration for the Li layers were considered. In one configuration, 
Li atoms in both layers 
sit above the hollow sites of the graphene layer which are  
labeled $\alpha$ (see Fig. \ref{fig:abc}). The stackings for C$_6$-Li-C$_6$-Li and C$_6$-Li-C$_6$-Li-C$_6$ are 
$A\alpha A\alpha$ and $A\alpha A\alpha A$, respectively. In this configuration,
C$_6$-Li-C$_6$-Li and C$_6$-Li-C$_6$-Li-C$_6$ will be simply denoted as 
C$_6$Li$_{\alpha\alpha}$ and C$_6$Li$_{\alpha\alpha}$C$_6$, respectively.
In the other configuration, whereas the Li atoms between the first and second graphene layers 
occupy positions labeled $\alpha$, the Li atoms above the second carbon layer occupy 
positions labeled $\beta$. In this case, the two Li layers are not directly above each other; 
the second layer is shifted, relative to the first, by a lattice vector of the graphene sheet. 
In this configuration, the stacking sequence  is $A\alpha A\beta$ for the Li-covered system,
and $A\alpha A\beta A$ for the intercalated system. In this configuration, the Li-covered
system will be denoted as C$_6$Li$_{\alpha\beta}$, while the intercalated system will be
written as C$_6$Li$_{\alpha\beta}$C$_6$.


For structures with three Li layers, two stacking configurations were considered. 
The first, in which the three Li layers are directly above each other, is 
 $A\alpha A\alpha A\alpha$. The corresponding system,  C$_6$-Li-C$_6$-Li-C$_6$-Li,
will be denoted as C$_6$Li$_{\alpha\alpha\alpha}$.
In the second configuration, the Li atoms between the first and second graphene layers 
occupy the hollow sites labeled $\alpha$, those between the second and third graphene 
layers sit in hollow sites labeled $\beta$, and those above the third graphene layer 
occupy sites labeled $\gamma$. The third layer of Li adatoms is shifted, 
relative to the second Li layer, by a lattice vector of the graphene sheet.
The stacking is of the form 
$A\alpha A\beta A\gamma$, and the Li-covered system is denoted by C$_6$Li$_{\alpha\beta\gamma}$. 

\section{Methods}
\label{sec:meths}

The reported total energy and electronic structure were obtained from 
 density functional theory (DFT) calculations using the all-electron, 
full-potential, linear augmented plane wave (FP-LAPW) method as implemented in the 
WIEN2K code \cite{WIEN2K}. The exchange-correlation potential was calculated using 
the generalized gradient approximation (GGA) as proposed by Perdew, Burke, 
and Ernzerhof \cite{PBE}. The radii of the muffin-tin spheres for the carbon and 
lithium atoms were taken as 1.36$a_0$ and 2.21$a_0$, respectively, 
where $a_0$ is the Bohr radius. The product R$_{MT}$K$_{max}$ was set to $7$, 
where R$_{MT}$ is the smallest muffin-tin radius and K$_{max}$ is a cutoff wave vector. 
The valence electrons wave functions inside the muffin-tin spheres were expanded in terms 
of spherical harmonics up to $l_{max}=10$, and in terms of plane waves with a 
cutoff wave vector K$_{max}$ in the interstitial region. The charge density was 
Fourier expanded up to a maximum wave vector G$_{max}=13a_0^{-1}$. 
Convergence of the self-consistent field calculations was attained with a total 
energy convergence tolerance of $0.1\ mRy$ and a total charge convergence tolerance of $0.0001\ e$.

The frequencies of the vibrational modes and the electron-phonon coupling parameter 
were calculated using the QUANTUM-ESPRESSO\cite{QE} package with norm-conserving pseudopotentials. 
The valence electrons wave functions and the charge density were expanded in 
plane waves using $40\ Ry$ and $300\ Ry$ energy cutoffs, respectively. The phonon frequencies 
were calculated within the linear-response framework on an $8 \times 8 \times 1$ 
phonon-momentum mesh with a $16 \times 16 \times 1$ uniform electron-momentum mesh. 
For the calculation of the electronic density of states and the electron-phonon coupling
strength, a uniform $40 \times 40 \times 1$ $\mathbf{k}$-point mesh in the Brillouin zone was used.

The Eliashberg spectral function $\alpha^2F(\omega)$ is given by
\begin{equation}
 \label{eqn:eliashberg}
\alpha^2F(\omega)=\frac{1}{N(0)N_kN_q}
\sum_{n\mathbf{k},m\mathbf{q},\nu}|g^\nu_{n\mathbf{k},m\mathbf{k}+
\mathbf{q}}|^2 \times \delta(\varepsilon_{n\mathbf{k}})\delta(\varepsilon_{m\mathbf{k}+\mathbf{q}})
\delta(\hbar\omega-\hbar\omega^\nu_{\mathbf{q}})
\end{equation}
where $N(0)$ is the total density of states at the Fermi level, and $N_k$ and $N_q$ are the total 
number of $\mathbf{k}$- and $\mathbf{q}$-points, respectively. In the above equation, 
$\varepsilon_{n\kv}$ is the energy, measured from the Fermi level, of an electron with
wave vector $\kv$ in energy band $n$,  $\omega_\qv^\nu$ is the frequency of a phonon with
wave vector $\qv$ and branch index $\nu$, and $g^\nu_{n\mathbf{k},m\mathbf{k}+\mathbf{q}}$
is the matrix element of the electron-phonon interaction.

The coupling strength of electrons to all phonon modes with frequencies below $\omega$ is 
defined as
\begin{equation}
 \label{lambda}
\lambda(\omega)=2\int_0^\omega\frac{\alpha^2F(\omega')}{\omega'}\ d\omega'
\end{equation}
The total electron-phonon coupling is calculated as $\lambda(\omega\to\infty)$. 
The values of the superconducting transition temperature were computed using the 
Allen-Dynes \cite{Allen-Dynes} modification of the McMillan formula \cite{Mcmillan},
\begin{equation}
 \label{allendynes}
kT_c=\frac{\hbar\omega_{log}}{1.2}exp\left[\frac{-1.04(1+\lambda)}{\lambda-\mu^*(1+0.62\lambda)}\right]
\end{equation}
where $\omega_{log}$ is the phonon logarithmic average frequency. 
The screened Coulomb pseudopotential \cite{Morel} $\mu^*$ was taken as $0.115$.


\section{Electronic Structure}
\label{sec:elec}

\begin{figure}[ht]
    \includegraphics{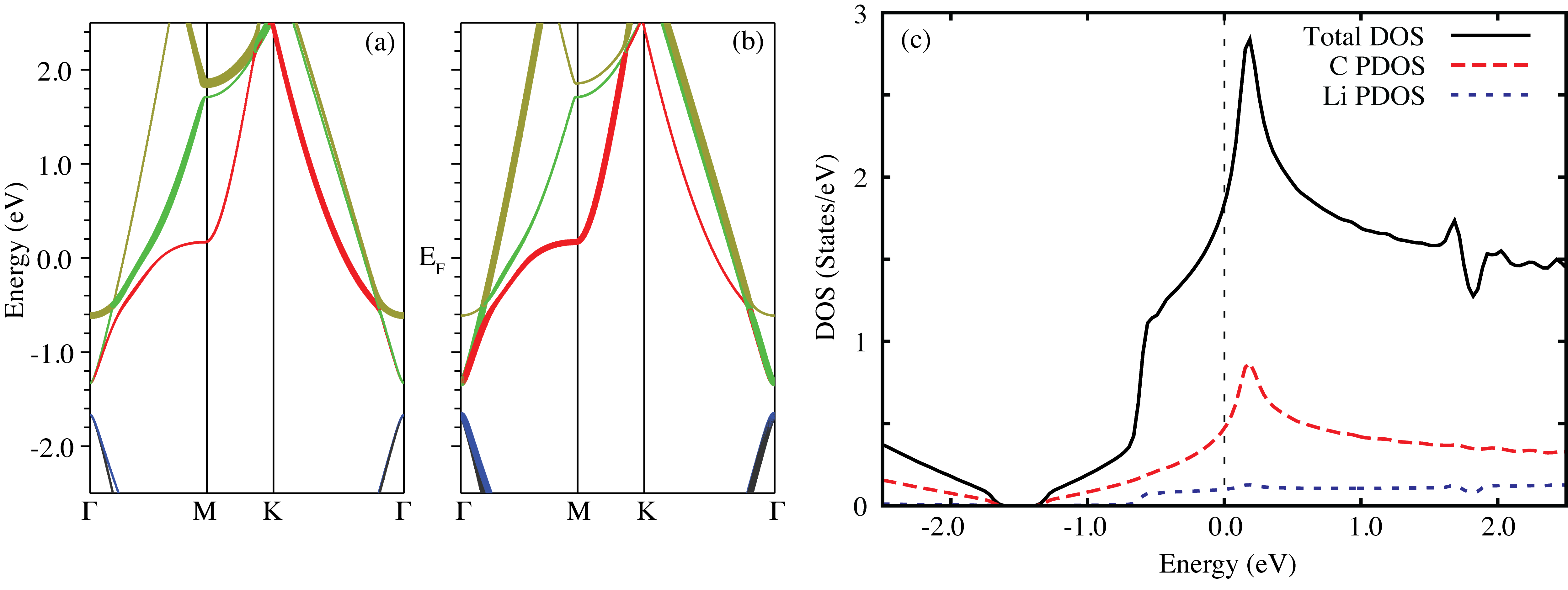}
\caption{(Color online) Plot of the energy bands ((a) and (b)) and density of states (c) in 
Li-covered graphene layer C$_6$Li along high symmetry directions in the Brillouin zone. 
The width of the bands at every $\mathbf{k}$-point is proportional to the contribution of 
(a) Li-derived states, and (b) C-derived states. The Fermi level is set at zero energy.}
\label{fig:elec-c6li}
\end{figure}

\begin{figure}[ht]
    \includegraphics{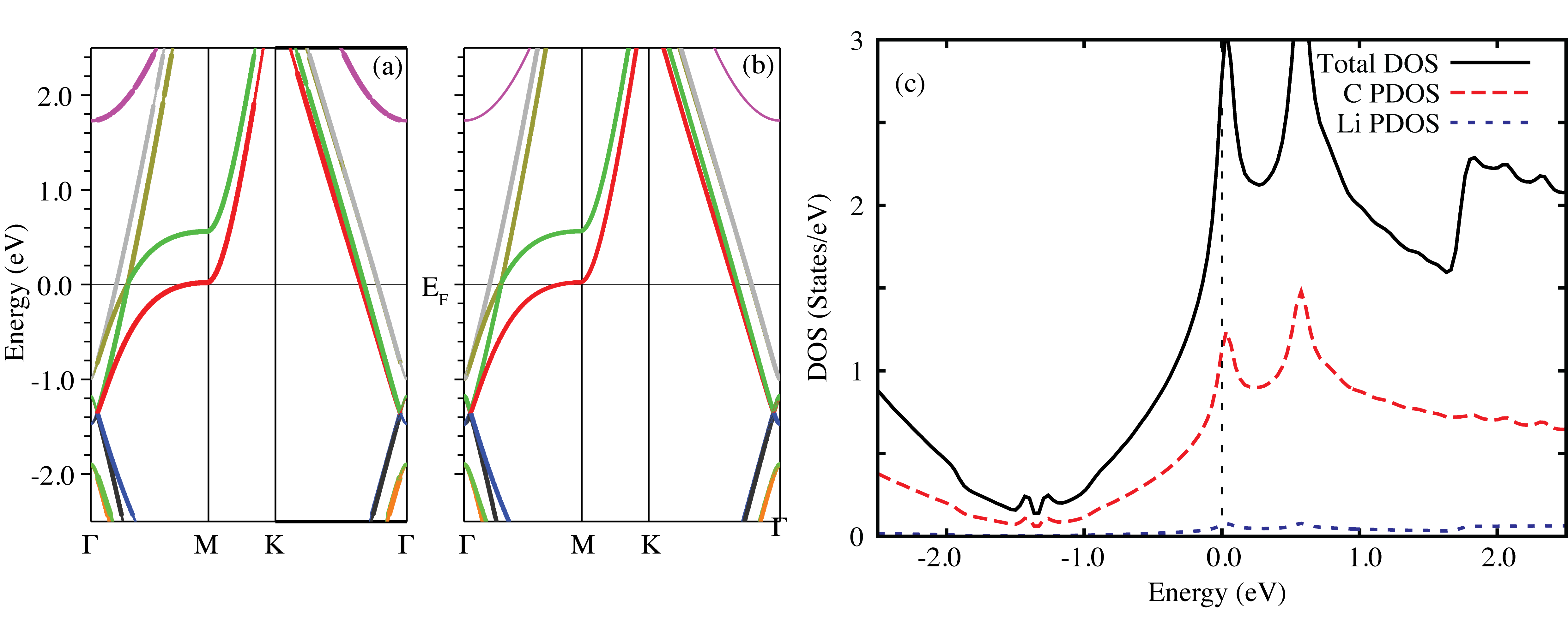}
\caption{(Color online) Plot of the energy bands ((a) and (b)) and density of states (c) in 
Li-intercalated graphene bilayer C$_6$LiC$_6$ along high symmetry directions in the Brillouin zone. 
The width of the bands at every $\mathbf{k}$-point is proportional to the contribution 
of (a) Li-derived states, and (b) C-derived states. The Fermi level is set at zero energy.}
\label{fig:elec-c6lic6}
\end{figure}

In Figures \ref{fig:elec-c6li} and \ref{fig:elec-c6lic6} we present the energy band 
dispersion and density of states (DOS) of lithium-covered graphene layer (C$_6$Li) 
and lithium-intercalated graphene bilayer (C$_6$LiC$_6$),
respectively. By inspection of the band-structure of C$_6$LiC$_6$, we notice that the 
interlayer band (derived from Li states) at the Fermi level is empty, as is the case 
in stage-1 Li-GIC, which is not a superconductor. In the case of C$_6$LiC$_6$, 
the $c$-direction confinement, imposed by the top and bottom graphene layers, keeps 
the interlayer band from being occupied. As Figure 3a reveals, the bottom of the 
Li-derived band is $\sim 1.75$ eV above the Fermi energy.
Based on these grounds, any further increase in the number of layers will not 
create new Li-derived states at the Fermi level. 

In the case of Li-covered graphene, we see that the Fermi level is crossed by an $s$-band 
derived from Li states. The removal of the top graphene layer has the effect of bringing down the energy of
the Li-derived band to the Fermi level; this suggests that there is a partial charge transfer from 
the Li atoms to the carbon sheet. The amount of charge transferred from each Li atom  
is calculated using Bader's quantum theory of ``atoms in molecules'' \cite{AIM}, and the 
results are presented in Table \ref{tab:aim}. The table reveals that there are two values 
of the charge transfer: 0.69 electrons (partial transfer) from an Li atom in the outer layer, 
and 0.88 electrons (almost complete transfer)
from an Li atom in an inner layer surrounded by two graphene layers. As in the case of GICs \cite{Csanyi}, 
the presence of an adatom band at the Fermi level increases the number of carriers, 
promotes electron coupling to the low energy carbon atoms out-of-plane vibrations, and generates coupling 
to the adatom vibrations. \cite{Calandra1}.

\begin{table}
\renewcommand{\baselinestretch}{1}
\caption{Charge transfer ($e^-$ per Li atom) from lithium atoms to graphene planes calculated 
using Bader's ``atoms in molecules'' (AIM) method for selected systems. For the Li-covered 
graphene systems with more than one lithium layer, the reported charge transfer is from Li 
atoms in the outer layer.}
\centering 
\begin{tabular}{c c |c c }
\hline
  Structure                           &  Charge Transfer ($e^-$/Li)   &  Structure                            &  Charge Transfer ($e^-$/Li) \\
\hline
C$_6$Li                               &  0.6847                                & C$_6$Li$_{\alpha\beta}$               &  0.6931 \\    
C$_6$Li$_{\alpha\alpha}$              &  0.6922                                & C$_6$Li$_{\alpha\beta\gamma}$         &  0.6892 \\ 
C$_6$Li$_{\alpha\alpha\alpha}$        &  0.6898                                & C$_6$LiC$_{6}$                        &  0.8784 \\ 
C$_6$Li$_{\alpha\alpha\alpha\alpha}$  &  0.6894                                & C$_6$Li$_{\alpha\beta}$C$_{6}$        &  0.8803 \\
\hline
\end{tabular} 
\label{tab:aim} 
\end{table}

In Figure \ref{fig:elec-c6li}(c) the partial density of states (PDOS) and the total DOS of C$_6$Li 
are compared. The total density of states is the sum of the atomic densities of states and 
the density of states in the interstitial region. It is important to note that the Li-projected 
DOS comes predominantly from the $s$-band. The LAPW method projects the DOS onto the 
muffin-tin spheres, yielding the atomic densities of states; the rest of the total DOS 
is assigned to the interstitial region. Figure 2c shows that a substantial portion
of the DOS results from the interstitial region. Since the interstitial space in this structure is 
mainly the space between the Li muffin-tin spheres, the contribution
of the Li-derived $s$-band to the total DOS is comparable to that from the graphitic $\pi$ bands.
 
We now focus on Li-covered and Li-intercalated graphene systems with two and three lithium layers. 
As discussed before (Sec. \ref{sec:systems}), we considered two stacking configurations when 
dealing with more than one lithium layer.

The first issue to address is to determine which stacking configuration yields a more stable structure. 
To study this, we carried out total energy calculations for the Li-covered and the  
Li-intercalated systems considered in this work. In all cases, we found that the lowest energy is obtained 
when the Li atoms of different layers sit directly above each other, so the Li layers 
are not shifted with respect to each other. In the case of Li-covered graphene layers, 
the structure C$_{6}$Li$_{\alpha\alpha}$ is lower in energy than its counterpart 
C$_6$Li$_{\alpha\beta}$ by 19.45 $meV$ per unit cell, the energy of 
C$_{6}$Li$_{\alpha\alpha\alpha}$ is lower than  C$_{6}$Li$_{\alpha\beta\gamma}$ by 
41.95 $meV$ per unit cell. In the case of Li-intercalated graphene layers, the structure 
C$_{6}$Li$_{\alpha\alpha}$C$_6$ is lower in energy than 
C$_{6}$Li$_{\alpha\beta}$C$_6$ by 20.42 $meV$ per unit cell.

Although it is concluded that it is energetically favorable for the lithium atoms of different 
layers to sit in equivalent hollow sites (Li stacking $\alpha\alpha\alpha$), there is some 
experimental evidence (Ref. \cite{Dresselhaus} and references therein) that at low temperatures 
the lithium intercalant layers in stage-1 LiC$_6$ GIC are sequentially arranged as $\alpha\beta\gamma$. 
For this reason we continue the comparative study of all structures 
regardless of their stacking configuration.

\begin{figure}[ht]
    \includegraphics{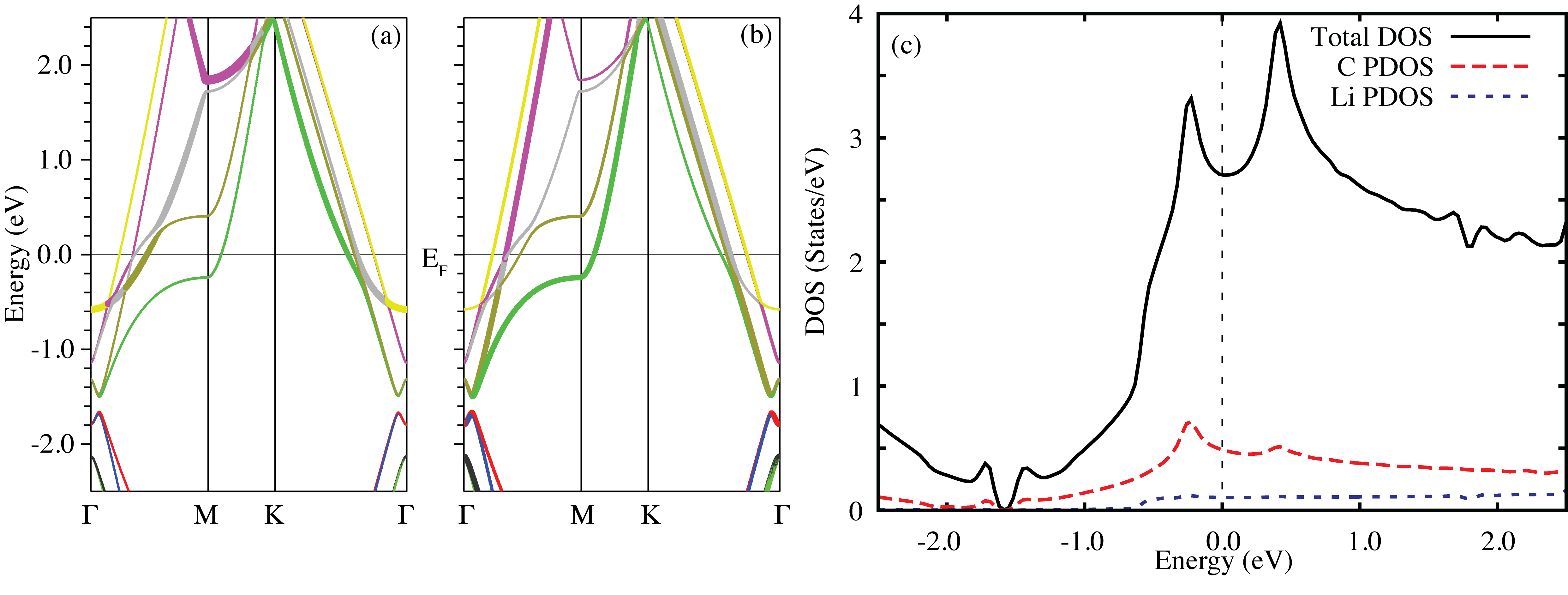}
\caption{(Color online) Plot of the energy bands ((a) and (b)) and density of states (c)  
in Li-covered graphene C$_{6}$Li$_{\alpha\beta}$ along a high symmetry path in the Brillouin zone. 
The width of the bands in (a) is proportional to the the contribution of the orbitals of 
the outer Li layer, while in (b) it is proportional to the contribution of the orbitals of the 
carbon atoms in the graphene layer adjacent to the outside Li layer. 
The Fermi level is set at zero energy.}
\label{fig:elec-c6liab}
\end{figure}

Figure \ref{fig:elec-c6liab} shows the calculated band structure and DOS of C$_6$Li$_{\alpha\beta}$. 
In similarity to the the C$_6$Li monolayer case, there is a lithium $s$-band
as well as graphitic $\pi$-bands crossing the Fermi level simultaneously. We note that
the $s$-band is derived from orbitals of the Li atoms lying in the outer Li layer. We also 
note that there is a substantial increase in the density of states at the Fermi level 
in comparison with the C$_6$Li case. In the  C$_6$Li$_{\alpha\beta}$ unit cell, 
there are two inequivalent Li atoms and two inequivalent C atoms. The atomic densities
of states shown in Fig. \ref{fig:elec-c6liab} are for the outer layer Li atoms and the C 
atoms in the adjacent layer.




We have also calculated the electronic structure of C$_6$Li$_{\alpha\beta\gamma}$ 
(plots of the energy bands and DOS for this structure are provided in the supplementary material). 
As expected from the previous case (C$_6$Li$_{\alpha\beta}$), in addition to the graphitic
$\pi$-bands, the Fermi level is crossed only by 
the band derived from the orbitals of the atoms of the outer lithium layer. We note 
the existence of a substantial increase in the DOS at the Fermi level relative to C$_6$Li, 
C$_6$Li$_{\alpha\alpha}$, and C$_6$Li$_{\alpha\beta}$. Analysis of the PDOS, projected on the 
carbon  states and lithium states, reveal that the main contribution to the total DOS 
results mainly from the carbon muffin-tin and interstitial regions. 

 
In the case of the Li-intercalated system denoted by C$_6$Li$_{\alpha}$C$_6$Li$_\beta$C$_6$, 
the interlayer bands are not occupied, and superconductivity is suppressed in this material, 
as we will demonstrate in Section \ref{sec:mode} by the computating the electron-phonon 
coupling and estimating the superconducting transition temperature. The energy bands and DOS
for this system are given in the supplementary material.


\section{Vibrational Modes and Electron-Phonon Coupling}
\label{sec:mode}
In Section \ref{sec:elec} we found that only Li-covered graphene systems exhibit a coexistence of 
graphitic $\pi$-states and lithium-derived $s$-states at the Fermi level. We also noted that 
$c$-direction confinement produced by the graphene layers push the bands derived from inner 
lithium atoms to energies well above the Fermi level. This means that only one Li-derived band 
crosses the Fermi level regardless of the number of Li-covered graphene layers. In addition, 
the total density of states at the Fermi energy in Li-covered graphene systems increases 
as the number of lithium layers increase,
a fact that could be relevant for the enhancement of the electron-phonon coupling strength. 
However, when the number of Li layers is large, the Li-covered graphene system, with 
equal number of Li and graphene layers, begins to look like stage-1 Li GIC, which is not
a superconductor. We conclude that there is an optimal number of Li layers which leads 
to the highest possible $T_C$ in these systems.
To this end, we present in this section electron-phonon coupling calculations for 
various Li-covered graphene systems, as well as the Li-intercalated graphene bilayer. 

The phonon dispersion curves for Li-covered and Li-intercalated graphene systems can be divided 
into three regions: a high energy region ($\omega_{\mathbf{q}\nu}>830\ cm^{-1}$) composed of in-plane 
carbon-vibration (C$_{xy}$), an intermediate region ($430\ cm^{-1}<\omega_{\mathbf{q}\nu}<830\ cm^{-1}$) 
composed of out-of-plane carbon-vibration (C$_z$), and a low energy 
region ($\omega_{\mathbf{q}\nu}<430\ cm^{-1}$) mostly composed of in-plane (Li$_{xy}$) and 
out-of-plane (Li$_{z}$) vibrations. It turns out that the low energy branches related to 
Li$_{xy}$ and Li$_z$, and their mixture with C$_z$, make the greatest contribution to the 
electron-phonon coupling parameter.

\begin{figure}[ht]
    \includegraphics[scale=0.7]{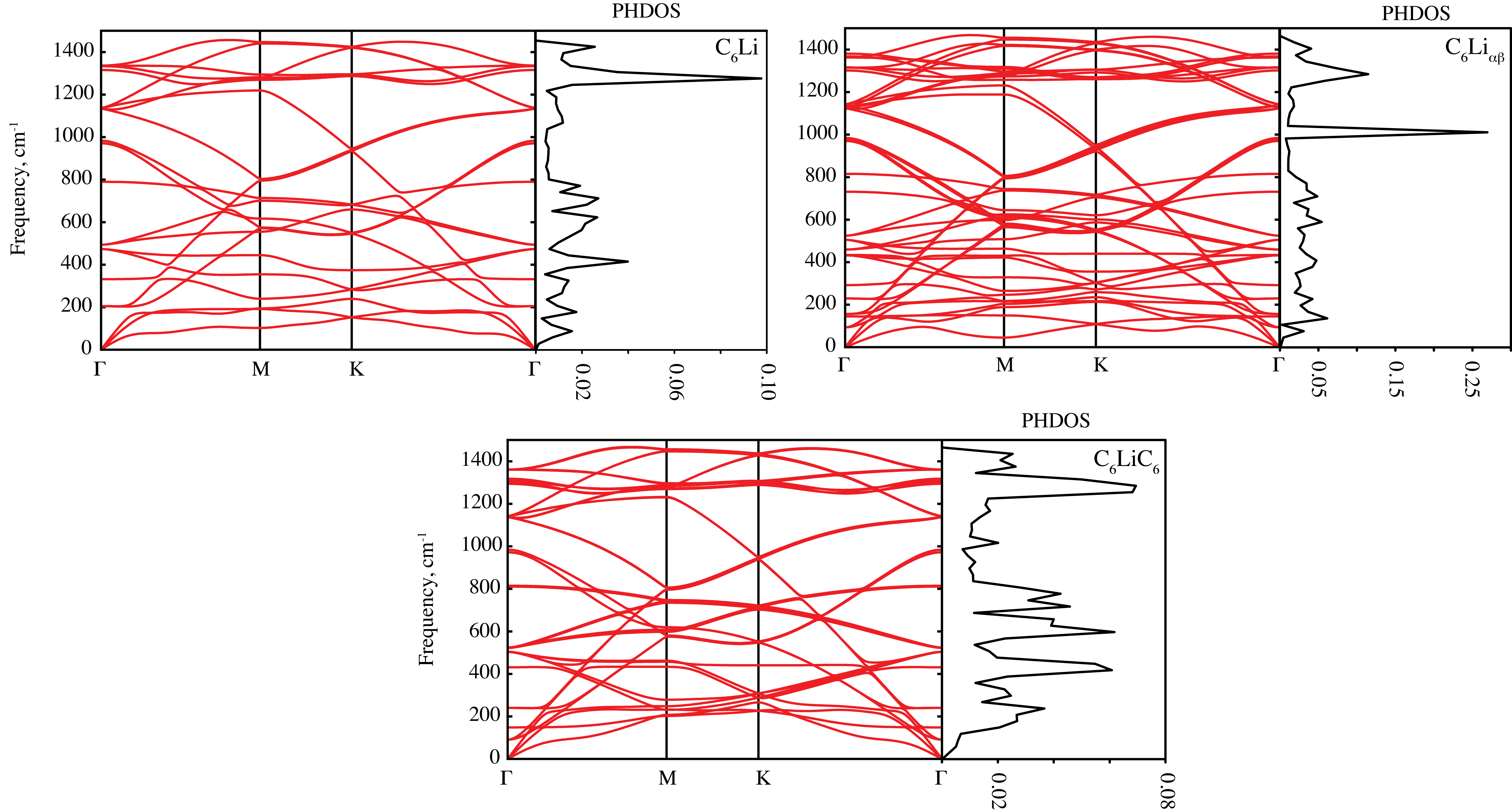}
\caption{(Color online) Phonon frequency dispersion and phonon density of states of
C$_6$Li (top left), C$_6$Li$_{\alpha\beta}$ (top right), and  C$_6$LiC$_6$ (bottom).}
\label{fig:phon-c6liall}
\end{figure}

In Figure \ref{fig:phon-c6liall} we present the phonon dispersion curves and the phonon 
density of states (PHDOS) in the Li-covered graphene systems C$_6$Li and C$_6$Li$_{\alpha\beta}$, 
and the Li-intercalated graphene bilayer C$_6$LiC$_6$. In the case of C$_6$LiC$_6$, the lowest 
energy branch 
is hardened compared C$_6$Li. 
Going from C$_6$Li to C$_6$Li$_{\alpha\beta}$, the lowest energy branch 
(corresponding to Li$_{xy}$-vibrations) in the $\Gamma K$ direction is softened, while branches 
immediately above (corresponding to Li$_z$) are hardened. We also note that the lowest energy 
branch in C$_6$Li$_{\alpha\beta\gamma}$ (shown in the supplementary material) is slightly hardened 
relative to C$_6$Li$_{\alpha\beta}$; this may explain the fact that the electron-phonon coupling is 
stronger in C$_6$Li$_{\alpha\beta}$ than in C$_6$Li$_{\alpha\beta\gamma}$, as we will see shortly. 
As a final note, we do not observe any significant effect on the high energy branches due to the 
increasing number of Li layers in Li-covered graphene systems.

\begin{figure}[ht]
    \includegraphics{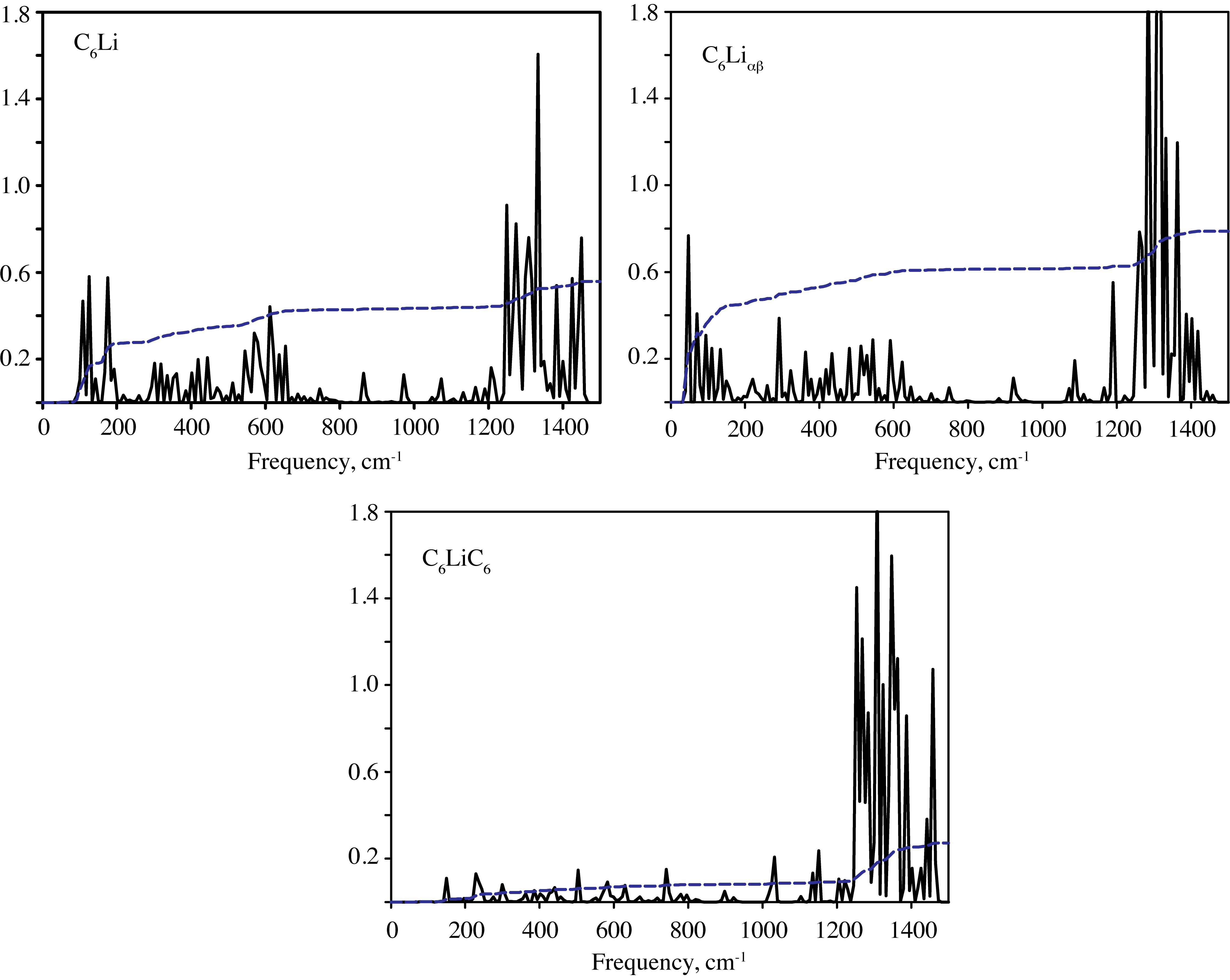}
\caption{(Color online) Eliashberg function $\alpha^2F(\omega)$ (continuous black line) and integrated 
electron-phonon coupling parameter $\lambda(\omega)$ (blue dashed line) for 
C$_6$Li (top left), C$_6$Li$_{\alpha\beta}$ (top right), and C$_6$LiC$_6$ (bottom).}
\label{fig:alphaFc6liall}
\end{figure}

The Eliashberg function $\alpha^2F(\omega)$, plotted in Figure \ref{fig:alphaFc6liall}, 
gives the contribution of each frequency to the total electron-phonon coupling constant $\lambda$ 
(see Sec. \ref{sec:meths}). In all cases we observe a considerable contribution to the total 
electron-phonon coupling from the carbon-carbon stretching modes, which are high in energy 
($\omega_{\mathbf{q}\nu}>1000\ cm^{-1}$). More importantly, a substantial contribution to the 
electron-phonon coupling from the Li modes and C out-of-plane vibrational modes is observed 
in all the Li-covered graphene systems. However, in the case of C$_6$Li$_{\alpha\beta}$ this 
contribution is significantly larger than that in C$_6$Li,  C$_6$Li$_{\alpha\alpha}$, 
and C$_6$Li$_{\alpha\beta\gamma}$ (the last two are shown in the supplementary material). 
For completeness' sake, we also show the Eliashberg function for C$_6$LiC$_6$ where we notice a 
weak coupling of electrons to the Li modes and C out-of-plane modes. This weak contribution in 
the Li-intercalated graphene bilayer is detrimental to the overall electron-phonon coupling 
parameter $\lambda$, which was calculated to be 0.29.

The calculated total electron-phonon coupling parameter $\lambda$ for  C$_6$Li is 0.55 with an estimated 
T$_c$ of 6.67 K. In the case of  C$_6$Li$_{\alpha\alpha}$ we calculated $\lambda=$0.65 and 
T$_c$=10.88K, for  C$_6$Li$_{\alpha\beta}$ we obtained $\lambda=$0.86 and T$_c$=13.54 K, and 
for C$_6$Li$_{\alpha\beta\gamma}$ we found that $\lambda$ is 0.57 and the estimated T$_c$ is 8.41 K. 
The results of the electron-phonon coupling calculations are summarized in Table \ref{tab:lambda}.

\begin{table}[ht]
\caption{{Summary of the electron-phonon coupling parameter ($\lambda$), logarithmic average 
frequency ($\omega_{log}$) and $T_c$ for all the studied systems.}}
\centering 
\begin{tabular}{c c c c||c c c c}
\hline
  Structure                           &  $\lambda$   &  $\omega_{log}$ ($cm^{-1}$)   &  T$_c$ (K) & Structure                           &  $\lambda$   &  $\omega_{log}$ ($cm^{-1}$)   &  T$_c$ (K) \\
\hline
C$_6$Li                               &  0.55        & 320.4                         & 6.67 & C$_6$Li$_{\alpha\beta}$               &  0.86        & 194.3                         & 13.54\\    
C$_6$Li$_{\alpha\alpha}$              &  0.65        & 308.2                         & 10.88 & C$_6$Li$_{\alpha\beta\gamma}$         &  0.57        & 357.8                         & 8.41\\ 
C$_6$Li$_{\alpha\alpha\alpha}$        &  0.56        & 296.7                         & 6.60 & C$_6$Li$_{\alpha\beta\gamma\alpha}$   &  0.46        & 469.0                         & 3.22\\ 
C$_6$Li$_{\alpha\alpha\alpha\alpha}$  &  0.49        & 373.4                         & 4.56 & C$_6$LiC$_{6}$ & 0.29 & 898.4 & 0.171\\
\hline
\end{tabular} 
\label{tab:lambda} 
\end{table}

\section{Conclusions}
\label{sec:con}
In this work, we presented the results of calculations of the electronic structure and 
electron-phonon interaction for several configurations of graphene-lithium systems. We 
assessed the structural stability and the strength of the electron-phonon coupling with 
respect to the stacking sequence and number of layers of lithium adatoms and the graphene sheets. 
We found that in all studied cases, the structures were stable, but the ones that yield the lowest 
energy were those where the Li adatom would sit in equivalent hollow sites; that is, with Li 
stacking of the form $\alpha\alpha$, $\alpha\alpha\alpha$,  and $\alpha\alpha\alpha\alpha$. 
Structural stability of the systems was later confirmed with the calculation of the phonon 
dispersion curves, which do not show the presence of any imaginary frequencies. We found that, 
in contrast to the corresponding GICs, Li-covered graphene systems exhibit an incomplete charge 
transfer from the lithium atoms to the graphene planes, giving rise to a lithium-derived band
which crosses the Fermi level simultaneously with graphitic $\pi$-bands. This enhances the 
electron-phonon coupling strength, which was found to be highest for the structure 
C$_6$Li$_{\alpha\beta}$. The calculated $\lambda$ for this structure was found to be $0.86$ 
with an average logarithmic frequency  $\omega_{log}=194.3$ cm$^{-1}$ and a critical 
superconducting transition temperature  T$_c=13.54$ K. The optimal value of T$_C$ is thus 
obtained for a system consisting of two graphene layers and two lithium layers.


One of the authors (RAJ) gratefully acknowledges support by NSF under grant No. HRD-0932421.
DMG would like to thank the CSULA-MORE programs and CSU-LSAMP fellowship for their support 
under grant No. HRD-1026102-518741.



\end{document}